\documentclass[english]{article}
\usepackage[T1]{fontenc}
\usepackage[latin9]{inputenc}
\usepackage{amsmath}
\usepackage{amsthm}
\usepackage{amssymb}
\usepackage{graphicx}

\makeatletter
\numberwithin{figure}{section}
\newcommand{\lyxaddress}[1]{
\par {\raggedright #1
\vspace{1.4em}
\noindent\par}
}

\makeatother

\usepackage{babel}
\begin{document}

\title{\textbf{How to distinguish an actual astrophysical magnetized black
hole mimicker from a true (theoretical) black hole}}
\maketitle
\begin{center}
\textbf{\large{}$^{1}$Abhas Mitra, $^{2}$Christian Corda and $^{3}$Herman
J. Mosquera Cuesta\medskip{}
}{\large\par}
\par\end{center}

\lyxaddress{$^{1}$\textbf{Homi Bhabha National Institute, Anushaktinagar Mumbai
400094, India, e-mail: abhasmitra@gmail.com}}

\lyxaddress{$^{2}$\textbf{International Institute for Applicable Mathematics
and Information Sciences, B. M. Birla Science Centre, Adarshnagar,
Hyderabad 500063 (India) and Istituto Livi - Via Antonio Marini, 9, I-59100 Prato, (Italy); e-mail: cordac.galilei@gmail.com}}

\lyxaddress{\textbf{$^{3}$Visiting Scientist of COLCIENCIAS, Programa Nacional
de Ciencias Basicas, Ciencia Espacial, Avenida Calle 26 No. 57 - 83,
Torre 8: Pisos 2-6, Cadigo postal: 111321 Bogot�, Colombia, e-mail:
herman.paye@gmail.com}}
\begin{abstract}
We remind that the ``ring down'' features observed in the LIGO gravitational
waves (GW) resulted from trembling of ``photon spheres'' ($R_{photon}=3M,\:by\:using\:units\:G=c=1$)
of newly formed compact objects and not from the trembling of their
event horizons (EH) $R=2M$ \cite{key-1}. Further, the tentative
evidence for late time ``echoes'' in GWs might be signatures of
horizonless compact objects rather than vacuum black holes (BHs) possessing
EHs. In general, in the past, many authors have considered the possibility
that the so-called BHs might be only BH mimickers (BHMs) having physical
surfaces ($R\approx2M$). Similarly, even for an ideal BH, the radius
of its shadow, which is $R_{shadow}=\sqrt{3}R_{photon},$ is actually
the gravitationally lensed shadow of its photon sphere. Accordingly,
any compact object having $R\le R_{photon}=3M$ would generate similar
shadow. Thus, no observation has ever detected any EH (any exact BH).
Also, by definition, it is fundamentally impossible to directly detect
any EH \cite{key-2}. One notes that all astrophysical compact objects,
except exact (chargeless) BHs, possess intrinsic magnetic moment that
is dominated by the dipole component. Even rapidly spinning neutron
stars (NS) are treated as spinning magnetic dipoles by ignoring the
weak additional multipole moments. Hence, since an exact BH has no
magnetic moment, a collapsing massive star must radiate away its multipole
magnetic moments one by one and left with mostly the dipole moment
immediately before becoming a BH \cite{key-3}. One also notes that
the magnetic field embedded in the accreting plasma close to the compact
object is expected to have a radial pattern of $B\sim r^{-1},$ while
the stronger BHM dipole magnetic field should fall off as $B\sim r^{-3}$.
Accordingly, it has been suggested that one may try to infer the true
nature of the so-called astrophysical BHs by studying the radial pattern
of the magnetic field in their vicinity \cite{key-4}. But here we
highlight that, close to the surface of BHMs, the magnetic field pattern
differs significantly from the same for non-relativistic dipoles.
In particular, we point out that, for ultra-compact BHMs, the polar
field is \emph{weaker} than the equatorial field by an extremely large
factor of $\sim\frac{z_{s}}{\ln z_{s}}$, where $z_{s}\gg1$ is the
surface gravitational redshift. We suggest that, by studying the radial
variation as well as significant angular asymmetry of magnetic field
structure near the compact object, future observations might differentiate
a theoretical BH from a astrophysical BHM. This study also shows that
even if some BHMs would be hypothesized to possess magnetic fields
even stronger than that of magnetars, in certain cases, they may effectively
behave as atoll type neutron stars possessing extremely low magnetic
fields.
\end{abstract}
\begin{quote}

\end{quote}

\section{Introduction}

While most of the astronomers believe that the astrophysical BH candidates
(BHCs) found in innumerable X-ray binaries and Active Galactic Nuclei
(AGN) are true mathematical BHs possessing EHs and singularities,
from time to time, in order to avoid many puzzles and paradoxes with
BH singularities and event horizons, many general relativists and
astrophysicists have suggested that such objects could be BHMs, which
are almost as compact as BHs ($R\approx R_{s}=2M$) but non-singular
and without exact EHs. The radius of a BHM may be expressed as 
\begin{equation}
R=(1+\epsilon)~2M;~~\epsilon\ll1,\label{eq: 1}
\end{equation}
whereas, for an exact BH, one has $e=0$ and $R=2M$. Here $R$ and
$R_{s}$ represent the (areal) radius of the compact object and the
Schwarzschild radius respectively (units with $G=c=1$). Technically,
one basic difference between a true BH and a BHM can be expressed
through the concept of gravitational redshift around the compact objects: 

\begin{equation}
z=\left(1-\frac{2M}{r}\right)^{-\frac{1}{2}}-1,\label{eq: 2}
\end{equation}
and on the surface of the compact object: 
\begin{equation}
z_{s}=\left(1-\frac{2M}{R}\right)^{-\frac{1}{2}}-1.\label{eq: 3}
\end{equation}
For a BHM, one finds 
\begin{equation}
z_{s}={\epsilon}^{-1/2}-1\approx{\epsilon}^{-1/2}.\label{eq: 4}
\end{equation}
For a mathematical true BH, one must have $z_{s}=\infty$ while for
a BHM, though $z_{s}$ is finite, it can be arbitrarily high: $1\ll z_{s}<\infty$.

It is important here to take note of a crucial aspect which is common
to both a true BH and its mimicker. Both of them reside within another
mathematical surface known as the \emph{photon sphere}, see Figure
1.1, that is situated at $R_{p}=3M$, and $z_{s}=z_{p}=\sqrt{3}-1\approx0.732$. 

\begin{figure}

\includegraphics[scale=0.25]{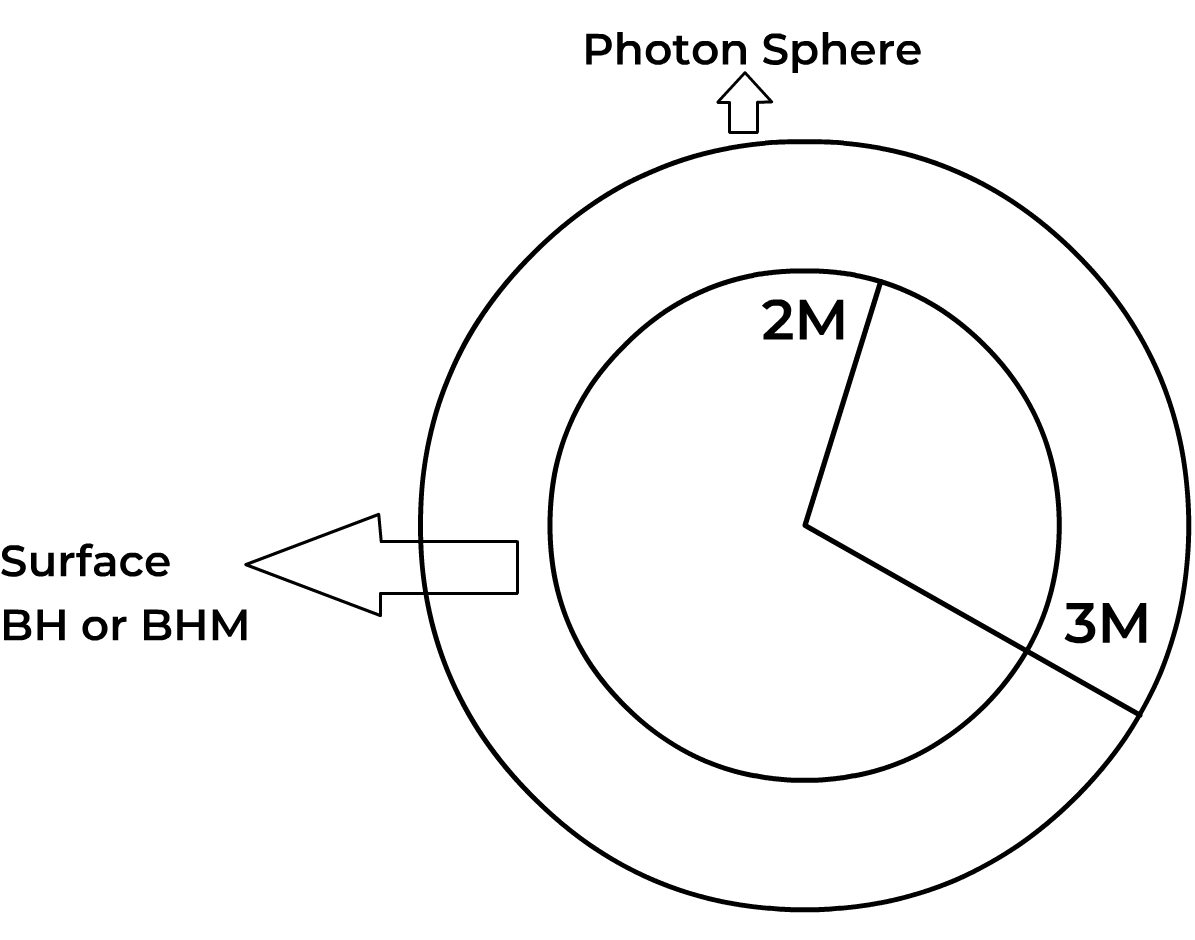}\caption{A picture of the same thing looking the other way!}

\end{figure}

In contrast, for a typical NS, one has $z_{s}\sim0.15$, and as if
the surface gravity of the photon sphere is 5 times stronger than
that of a NS. The surface gravity of the photon sphere is already
so strong that, \emph{on the photon sphere, photons can move in unstable
closed circular orbits} \cite{key-5}. In contrast, the circular orbits
of finite rest mass particles around a BH or a BHM end at $R=3,\;R_{s}=6M$,
and this defines the Innermost Stable Circular Orbit (ISCO). Within
the ISCO, for a non-magnetic BH or BHM, material particles tend to
accrete in radial directions and in near free fall.

Within the photon sphere, not only material particles, but even photons
tend to preferentially move inward, as the angle of the escape cone
shrinks rapidly. In order to escape from the interior of the photon
sphere, a photon must ejected in a perfectly radially outward direction.
And the probability of escape diminishes rapidly as $(1+z)^{-2}$
while $z\to\infty$ as $r\to2M$ \cite{key-5}. It is for this reason
that the photon sphere acts as a ``virtual EH'' for the BH and its
mimicker. Thus, the compact object looks almost ``black'' to a far
away astronomer.

At the beginning, we shall emphasize the fact that it is not possible
to detect any EH directly, verbatim from \cite{key-2}:

``Recently, several ways of verifying the existence of black hole
horizons have been proposed. We show here that most of these suggestions
are irrelevant to the problem of the horizon, at best they can rule
out the presence of conventional baryonic matter in the outer layers
of black hole candidates. More generally, we argue that it is fundamentally
impossible to detect in electromagnetic radiation direct evidence
for the presence of a black hole horizon. This applies also to future
observations, which would trace very accurately the details of the
space-time metric of a body suspected of being a black hole. Specific
solutions of Einsteins's equations lack an event horizon, and yet
are indistinguishable in their electromagnetic signature from Schwarzschild
black holes.''

In the following Section, we shall dwell on this issue in detail.
Next, we shall point out that studies carried out right from the era
of Ginzburg \cite{key-3} have suggested that the so-called BH candidates
might be ultra-magnetized BH candidates instead of exact BHs possessing
no magnetic fields. We shall also discuss that there are indeed evidences
that the astrophysical BH candidates may be possessing strong magnetic
fields. Then, we shall point out that there are many indirect and
observational evidences that the so-called astrophysical BHMs might
be magnetized BHMs.

In high-energy astrophysics, two kinds of X-ray binaries are of special
importance: 
\begin{enumerate}
\item Neutron Star X-ray Binaries (NSXBs); 
\item Black Hole Candidate X-ray Binaries (BHCXBs). 
\end{enumerate}
For understanding many astrophysical phenomena, such as launching
of relativistic jets from close to the compact objects, one requires
that the compact object must have reasonably strong intrinsic magnetic
field. Indeed, it is much easier to under the phenomenon of relativistic
jets from NSXBs like Cir X-1 \cite{key-7}, Sw J0243 \cite{key-8},
Aql X-1 \cite{key-9}, Her X-1 \cite{key-10} and others \cite{key-11,key-12}
containing spinning magnetized stars. Thus, for explaining the same
phenomenon from BHMXBs and AGN, supposed to be powered by supermassive
BHs, astrophysicists are compelled to assume that, in the presence
of exterior disk magnetic field, also spinning BHs should behave as
spinning conductors and develop induced electromagnetism. They justify
such an expectation on the plea that the exterior magnetic field lines
``thread'' the central BH \cite{key-13}.

However, also a spinning insulator sphere will be threaded by surrounding
magnetic field lines, and this does not mean that spinning insulating
sphere can ever behave as a spinning conducting sphere. We shall explain
why the magnetic field around accreting BHs could be only due to what
is embedded in the accreting plasma, and which may have a spatial
pattern of $B\sim r^{-1}$ \cite{key-4}. In contrast, the dominant
dipole magnetic field of a BHM is expected to a spatial variation
of $B\sim r^{-3}$ over and above the weaker magnetic field embedded
in the accreting plasma. Accordingly, Lobanov \cite{key-4} has suggested
that magnetized BHMs may be observationally differentiated from true
BHs by studing the pattern of radial variation of the magnetic field
around supposed astrophysical BHs. Here we shall highlight the fact
that, spatial variation apart, the angular pattern of the magnetic
field in close vicinity of the BH candidates could help in unravelling
their true nature. 

\section{Have Exact BHs Been Already Detected?}

Recently, two research fields of gravitation made important developments:
(i) GW detection from the formation of supposed BHs coalescence of
two compact objects \cite{key-14}; (ii) Imaging of the supposed supermassive
BH in the AGN M87 \cite{key-15}. Following these two astronomical
milestones, it is widely believed that astronomical observations have
finally pinned down on exact EHs, which, in turn, would prove the
existence of true mathematical BHs. For instance, the detection of
a final ``ring down'' phase from the maiden GW event GW150914 was
interpreted as the signature of trembling of a new formed deformed
EH, resulting from the merger of two massive compact objects \cite{key-14}.
But it was soon pointed out that the ringdown signal was the signature
of the trembling deformed photon sphere (or photon ring for spinning
axially symmetric compact objects) of the newly formed compact object,
rather than trembling of any true EH \cite{key-1}.

In comparison, the image/shadow of the supposed BH in the AGN M87,
detected by the Event Horizon Telescope (EHT) in 2019 \cite{key-15},
is considered the ultimate direct evidence of existence of EHs. But
an honest introspection will show that the released fuzzy image and
its central dark region is no evidence for any EH. There are two reasons
behind this assertion:
\begin{enumerate}
\item As already mentioned, no external observer can peer beneath the photon
sphere where light can move in closed circular orbits. Therefore,
to a distant astronomer the photon sphere appears as almost \textquotedbl{}black\textquotedbl{}
because hardly any radiation can pierce out of the clutch of its strong
gravity. Thus, at first sight, one may think that the central dark
patch in the image of a BH has a radius of $3M$ instead of $2M$
\cite{key-5}.
\item But even this is not true. Any light or radiation (say from the accretion
disk) coming from behind the photon sphere gets strongly gravitationally
lensed by the photon sphere. Then, in the ideal case, the shadow gets
inflated by a factor of $\sqrt{3}$. 
\end{enumerate}
Thus, for an idealized Schwarzschild BH or its mimicker, the radius
of the shadow is (see Figure 1.2) \cite{key-5,key-16}:

\begin{equation}
R_{shadow}=3\sqrt{3}M,~if~R\le3M,or~z_{s}\ge0.732.
\end{equation}
\begin{figure}
\includegraphics[scale=0.125]{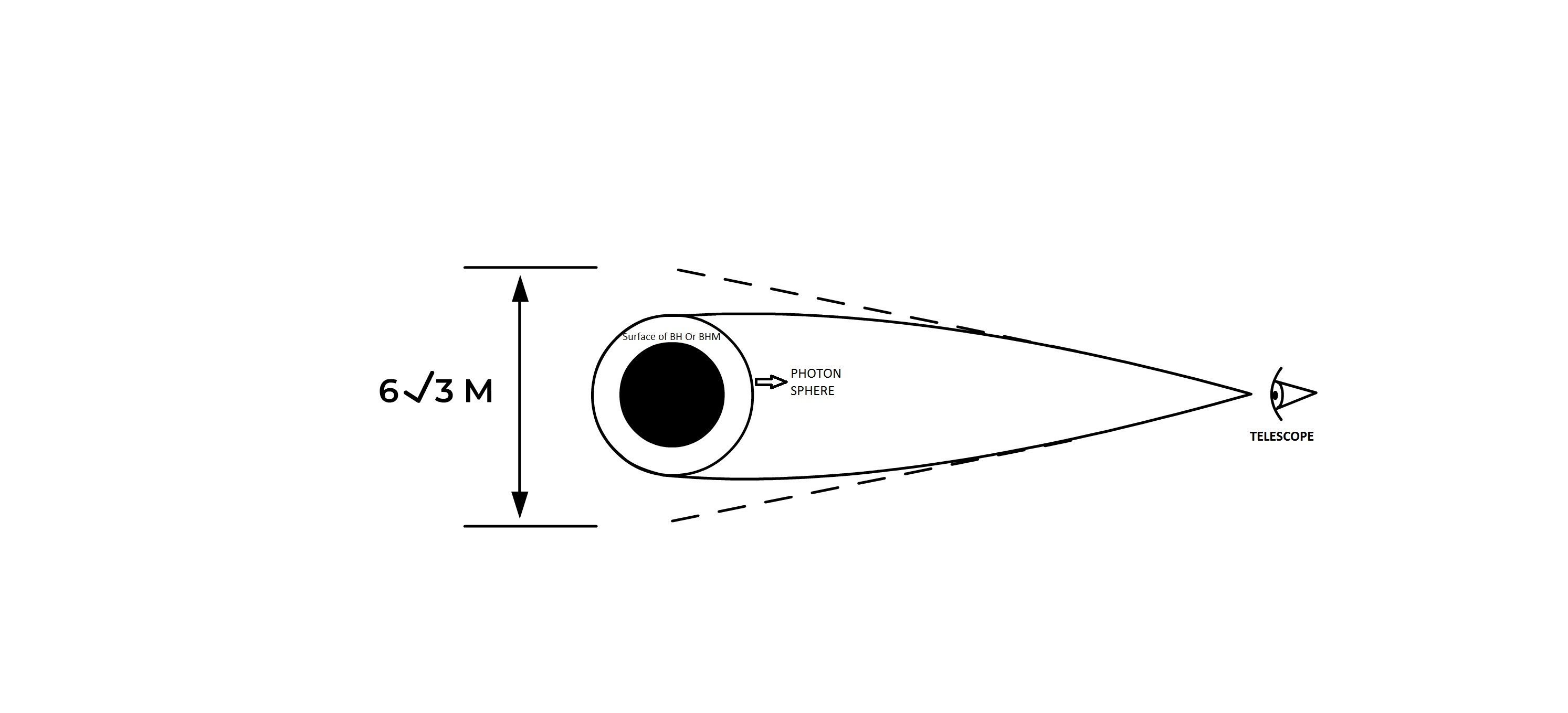}\caption{The shadow of an astrophysical compact object}
\end{figure}

Since it is the shadow of the photon sphere ($z_{p}\approx0.732$)
that the astronomer can detect, even a fat BHM having a radius just
smaller than the photon sphere, say $R=2.99M,$ will yield an exterior
shadow of the same size $R_{shadow}=3\sqrt{3}M$. In view of this
rather unexpected fact, no telescope can distinguish between: (i)
a true BH ($R=2M$) or (ii) an ideal BHM ($R\approx2M$), (iii) even
a crude fat BHM ($R\le3M$). The shadow size would be different from
these class of objects only when the radius of the compact object
would be larger than its photon sphere ($R>3M$) and $z_{s}<z_{p}\approx0.732$,
for instance, for a NS \cite{key-16}, one will have

\begin{equation}
R_{shadow}=(1+z_{s})R;~~ifR>3M,~or~z_{s}<0.732
\end{equation}
Thus, the shadow/image of a NS having a radius of $10\:km$ and $z_{s}=0.15$
will appear as a circle of radius $11.5\:km$. In contrast, for a
white dwarf having $z_{s}\ll1$, there will hardly be any enhancement
in the size of the shadow/image. 

\subsection{Tentative Evidence Against Exact Black Holes}

GWs generated from the trembling of the new born unsteady photon sphere
travel both outward and inward. The outward moving GWs may exhibit
as ring down waveform. If the underlying new born compact object would
be a true vacuum BH, it would gulp down all inward moving GWs. In
such a case, external telescopes would not detect any subsequent GW
emission for the given event. But, in that case, the underlying compact
object is a EH less BHM. The inward travelling GWs will initially
remain trapped between its surface and the photon sphere. However,
part of the trapped GWs would diffuse out later following repeated
reflections from the physical surface of the BHM. In such a case,
there might be subsequent GW echoes. There have been some weak evidences
for GW echoes associated with three cases of GW detection: GW150914,
GW151226, and LVT151012 \cite{key-17}. In comparison, there has been
stronger evidence for GW echoes for the binary neutron star merger
event GW170817 \cite{key-18}. Thus, tentative detection of GW late
echoes might suggest that the compact objects formed by the coalescence
of either so-called BHs or even NS might be BHMs having physical surfaces
and not vacuum BHs possessing EHs \cite{key-19}. 

One notes that the Sun is a magnetized ball of plasma and vulnerable
to unpredictable eruptions driven by locallized turbulence of magnetized
plasma, like flares and Coronal Mass Ejections (CME), which may also
be nicknamed as ``blurps'' or ``blasts''. In contrast, nothing
can escape a true BH, and neither can be strong magnetospheric instability
around magnetic fieldless BHs. Though there could be eruptions from
the accretion disk, the strength of eruptions are expected to be modest
as magnetic energy density in the far away disk would be much smaller
than say what is present in a magnetized NS. And such disk eruptions
should be evident for both NSXBs as well as BHCXBs. Thus, one would
expect magnetic field driven eruptions to be much stronger for NSXBs
whose disk and magnetosphere are dominated to be magnetic fields,
than the same for BHCXBs. But, in reality, astrophysicists find the
BHCXBs to be much more volatile and much more transient than NSXBs.
In fact, there are often news items of ``belching'' and ``burping''
from astrophysical BHs. Such belchings must be sudden nassive outflows
of plasma from close to the BHs. A recent case of discovery of such
``burps'' was from the AGN SDSS J1354+1327 \cite{key-20}. Such
outbursts, though not understood, are conjectured to resulting from
episodes of extremely high rate of accretions whose origin is unknown.
If we scale down to stellar mass compact objects, there should be
``burps'' from NSXBs and which should be stronger than outbursts
from BHCXBs. This is so because, in the absence of any magnetic resistance,
accretion onto BHs should be rather monotonous effect in contrast
to the same for magnetized NSs. But, in reality, BHCXBs are found
to be much more turbulent giving rise to transient events in comparison
to NSXBs. Such an apparent contradiction would be removed if BHs were
magnetized compact objects having magnetic field structures much more
complicated than typical (simple dipole) NS magnetic fields. 

As mentioned above, X-ray flares from BHs are very commonplace, and
surprisingly BHXBs are much more violent than their NS counterparts.
Here we recall one particular X-ray flare detected in September 2014
by NASA's Explorer missions Swift, and the Nuclear Spectroscopic Telescope
Array, or NuSTAR \cite{key-21}. For the first time, it was found
that the X-ray flare was triggered by a burst of corona from close
to the BHC weighing about two hundred million solar masses in the
AGN Markarian 335. It is difficult to understand how there could be
emergence of corona from the vicinity of a mathematical BH which has
no magnetic field, no matter and from which nothing can escape. 

\section{More Direct Evidence Against Magnetic Fieldless Black Holes}

Since astrophysical plasma is overall neutral, astrophysical (true)
BHs should be chargeless and devoid of any intrinsic magnetic field.
However, astrophysical BHs are expected to be surrounded by accretion
disks that possess magnetic fields, and, in turn, can generate some
magnetic fields in the vicinity of the BHs. This problem was studied
in detail by Shakura and Sunyaev long ago \cite{key-22}. They arrived
at two important conclusions:

(i) The magnetic fields around astrophysical BHs should be negligible
and \emph{not} compared to the value which may dynamically dominate
the plasma. This is the reason that

${\bullet}$ The radius of the inner accretion disk is determined
by purely general relativistic reasons at the ISCO $r_{i}=3R_{s}=6M$,
and not by any magnetic field effect.

${\bullet}$ Within the ISCO, the accretion flow around BHs is quasi-spherical
and eventually becomes perfectly spherical near the EH.

In contrast, for a strongly magnetized NS, the inner radius of the
accretion disk is determined by equating two opposing effects \cite{key-23,key-24}.
At the inner accretion disk it is:

Outward Pressure Due To NS Intrinsic Magnetic Field ($\frac{B^{2}}{8\pi}$)
= Inward Ram Pressure of Inflowing Material

Thus, by definition, the region interior to the accretion disk of
a strongly magnetized NS (even a white dwarf) is \emph{dynamically
dominated} by the intrinsic magnetic field, and the accretion plasma
usually gets guided by the dipole magnetic field lines of the NS towards
its polar regions which forms the X-ray hotspots.

On the other hand, for weakly magnetized recycled millisecond pulsars
possessing magnetic fields as low as $10^{8}\,G$ , the accretion
flow, in view of the ram pressure, may accrete in a quasi-spherical
manner, smeared all over the surface. Even in this case, the ambient
magnetic field may be dominated by the NS magnetic field.

(ii) Another conclusion of Shakura and Sunaev \cite{key-22} was that,
verbatim, ``Hence, within the disk, the field is most likely to be
\emph{chaotic and of small scale}''.

One notes here that the launching and subsequent propagation of ultra-relativistic
jets, often associated with a BH, require the existence of fairly
organized and large scale magnetic fields, that is difficult to explain
by invoking the weak, chaotic and small scale magnetic field of the
disks. Shakura and Sunayev \cite{key-22} found that, for an event
of highest rate mass accretion (${\dot{M}}$), at the inner edge of
the accretion disk ($r_{i}$), for a BH of $10M_{\odot}$, one should
have $B_{i}\ll10^{8}\,G.$ As $r_{i}=6M$ increases with the BH mass,
and the ram pressure decreases at larger $r_{i}$, one might tentatively
expect the inner disk magnetic field to fall of as $B_{i}\sim M^{-2}$.
Then, any magnetic field of accretion origin must decrease with low
mass accretion rate (${\dot{M}}$).

$\bullet$ However, in 2003, from polarimetric observations of the
X-ray binary Cygnus X-1, that apparently contains a BH of mass $M\sim15M_{\odot}$,
it was found that is $B_{i}\approx10^{8}\,G$ \cite{key-25}. In fact,
in 2009, Karitskaya et al. concluded that, verbatim, ``the so-called
BH in Cygnus X-1 has a dipole magnetic moment of $10^{30}\,Gcm^{3}$'',
and accordingly, they called it ``Magnetic Extremely Compact Object
(MECO)'' rather than true BH \cite{key-26}.

$\bullet$ There have been great strides in direct determination of
magnetic fields around astrophysical BHs, and, in 2013, a dynamically
strong magnetic field was discovered close to Sgr A$^{*}$, the supposed
supermassive BH at our galactic centre (GC) \cite{key-27}.

$\bullet$ In the following year, it was found that the jet-launching
regions of 76 radio-loud active galaxies are threaded by \emph{dynamically
important fields} \cite{key-28}.

$\bullet$ In 2015, by studying the interferometric observations at
$1.3-$millimeter wavelength that spatially resolve the linearly polarized
emission from the GC supermassive BH, Sagittarius A$^{*}$, Johnson
et al. \cite{key-29} found evidence for partially \emph{ordered magnetic
field} near the EH, on scales of $\sim6$ Schwarzschild radii.

$\bullet$ In the same year 2015, by studying polarization of millimeter
radio waves caused by Faraday rotation, Atacama Large Millimeter/submillimeter
Array (ALMA) detected the presence of \emph{organized magnetic field}
in the AGN PKS 1830-21, having strength of a few Gauss (or even higher)
at a distance of around $0.01\,pc$ from the supposed BH \cite{key-30}.
This is a clear indication of very high magnetic fields at the jet
base of the compact object \cite{key-30}. 

All such evidence of unexpectedly high and organized magnetic fields
around supposed BHs may be best explained if the pertinent compact
objects possess sufficiently strong organized large scale dipole intrinsic
magnetic fields. 

\section{Can Magnetic Fieldless Spinning Black Holes Act As Strongly Magnetized
Pulsars?}

It is well known that, when any conductor moves in an external magnetic
field, it develops an induced electromagnetism. If the conducting
sphere itself is magnetized, it can develop an induced electric field
even in the absence of any exterior magnetic field. This is the reason
that spinning magnetized NSs act as electromagnetic pulsars \cite{key-31}.
The energy radiated by pulsars originate from their immense rotational
kinetic energies. By taking this cue, in order to explain the source
of energy of ultra-relativistic jets associated with astrophysical
BHs, it was assumed that spinning BHs too may act like spinning NSs
or pulsars \cite{key-13}. The Authors of \cite{key-13} built on
this hypothesis by attendant mathematics. In particular, Blandford
and Znajek stressed that, verbatim, \cite{key-13}: ``When a rotating
black hole is threaded by magnetic field lines supported by external
currents flowing in an equatorial disc, an electric potential difference
will be induced.''

Let us introspect here the physical meaning of the term \textquotedbl{}threaded
by magnetic fields\textquotedbl{}. If any object moves in an exterior
magnetic field ${\vec{B}}$, by ignoring the effect of bound molecules,
the exterior magnetic field lines thread the object. Then, by simple
application of Special Theory of Relativity (STR), a magnetically
induced electric field crops up within its interior (by restoring
standard units):

\begin{equation}
{\vec{E}}_{magnetic}=\frac{\vec{v}\times{\vec{B}}}{c},
\end{equation}
if the velocity of the moving object
\begin{equation}
|{\vec{v}}| \ll c
\end{equation}
Such a magnetically induced electric field would be generated in the
interior of all moving insulator too, and say within a spinning mica
ball.

Thus, from a purely mathematical point of view, the mica ball too is threaded
by the exterior magnetic field and it develops an induced electric
field too. In other words, the mical ball develops induced electromagnetism
by virtue of its motion in an exterior magnetic field. But is this
conclusion physically correct? The answer is obviously \textquotedbl{}NO\textquotedbl{}.
An insulator cannot indeed develop any induced electromagnetism (by
neglecting effects of bound molecules like diamagnetism). 

In order to understand this obvious fact, we need to understand the
fundamental reason for induced electromagnetism in materials is the
action of magnetic part of the Lorentz Force: 
\begin{equation}
{\vec{F}}_{magnetic}=e{\vec{E}}_{magnetic}=e{\vec{v}}\times{\vec{B}}
\end{equation}
on the \emph{free electrons} inside the material. One notes that the
electrons and ions bound to atoms and molecules do not feel this induced
Lorentz force despite the \emph{mathematical} presence of magnetic
and frame transferred electric fields. A good conductor like Copper
has a free electron density of $N_{e}\sim8.5\times10^{28}$ electrons
$m^{-3}$ and a resistivity of only $1.7\times10^{-8}\,ohm\,m$. In
contrast, the insulator mica has a $N_{e}\sim10^{6}$ $m^{-3}$ and
a huge resistivity of around $10^{23}$ $ohm\,m$. Further, for a
perfect insulator, one has $N_{e}=0$, and accordingly, conductivity
$\sigma=0$ and resistivity $\infty$.

Then, for a perfect insulator, the induced current density reads

\begin{equation}
\vec{J}_{insulator}=\sigma~{\vec{E}}_{induced}=0~\times{\vec{E}}_{induced}=0.
\end{equation}
In contrast, for a perfect conductor, free electrons promptly \emph{redistribute
themselves to set up an opposing electric field }

\begin{equation}
{\vec{E}}_{reaction}=-{\vec{E}}_{induced}
\end{equation}
so that

\begin{equation}
\vec{E}_{net}^{conductor}={\vec{E}}_{reaction}+{\vec{E}}_{magnetic}=0.
\end{equation}
As a result, the net Lorentz force on the free electrons of the perfect
conductor vanishes: 
\begin{equation}
{\vec{F}}=e~\vec{E}_{net}^{conductor}=e~\times0=0.
\end{equation}
 In contrast, for an ideal insulator, one has free electron density
$N_{e}=0$ and resistivity $\infty$. Thus, \emph{no back reaction
electric field is set up in the absence of motion of free electrons}

\begin{equation}
\vec{E}_{net}^{insulator}={\vec{E}}_{magnetic}\neq0.
\end{equation}
But, despite this mathematical frame transformed electric field, there
is no flow of current within an insulator moving in an exterior magnetic
field: 

\begin{equation}
{\vec{J}}^{insulator}=\sigma~{\vec{E}_{net}^{insulator}}=0\times{\vec{E}_{net}^{insulator}}=0
\end{equation}
Hence,though  exterior magnetic field lines thread both a conductor and an
insulator,  it is only the former which \emph{gets organically connected
to the exterior magnetic field} and develops induced electromagnetism.
Therefore, it is important to appreciate the fact that \emph{it is
not mere threading of exterior magnetic field lines in a material,
but, on the other hand, the presence or absence of super dense free
electrons which determines whether the material will develop induced
electromagnetism or not}.

Further, though insulators possess tiny but negligible free electron
density, a perfect vacuum, by definition, has $N_{e}\equiv0$ and
$\sigma\equiv0$.

Nonetheless, an ultra-strong exterior magnetic field of the order
of $B>B_{cr}$ may generate induced electromagnetic properties by
virtue of Quantum Electrodynamics (QED) effects \cite{key-32}, where

\begin{equation}
B_{cr}\equiv\frac{m_{e}^{2}c^{3}}{\hbar|e|}=4.4\times10^{13}~{\rm G}.
\end{equation}
Here $m_{e}$ and $e$ are the mass and charge of an electron. Thus,
barring such a QED scenario, if we imagine a certain region of vacuum,
say a glass box with interior vacuum, to be moving in an exterior
magnetic field, the vacuum cannot develop even the negligible induced
electromagnetism that an insulator might generate.

Let us be cautious here about a related fact. We know that electromagnetic
waves can propagate unimpeded in a vacuum having an \emph{impedance}
of 

\begin{equation}
Z_{0}=\sqrt{\frac{|{\vec{E}}|}{|{\vec{H}}|}}=\mu_{0}~c=377\,{\rm Ohm},
\end{equation}
where ${\vec{E}}$ and ${\vec{H}}$ are the electric and magnetic
field strengths of the propagating electromagnetic field, and $\mu_{0}$
is the magnetic permeability of the vacuum. But we must \emph{not}
confuse this vacuum impedance with any finite resistivity or conductivity
of the vacuum. If vacuum possessed any conductivity, we would always
get strong electric shock from a current carrying conductor even without
touching it ever.

Incidentally, BHs too have the same impedance $Z_{0}=377\,Ohm$ ,
that only reiterates that mathematical BHs are vacuum solutions without
any matter except for their central singularities \cite{key-33}.
Therefore, from the viewpoint of basic physics, it is rather meaningless
to assume vacuum BHs, having zero free electrons and from whose singularity
neither light nor electric current can flow outward, can behave as
pulsars that are almost perfect conductors and possess strong magnetic
fields. 

One also notes that, for a non-vacuum medium, one has

\begin{equation}
{\vec{B}}=\mu_{0}({\vec{H}}+{\vec{M}}),
\end{equation}
where ${\vec{M}}$ is the magnetization density vector that encompasses
the effects like polarization of matter due to external magnetic fields.
But

\begin{equation}
{\vec{M}}=0;~~for~vacuum
\end{equation}
and ${\vec{B}}=\mu_{0}~{\vec{H}}$. The fact that, for both the vacuum
and the BH, one has the same impenance $Z_{0}$, reconfirms that classical
vacuum has no magnetic property, ${\vec{M}}=0,$ even though magnetic
field lines may thread it. 

Looking back to Goldreich and Julian model of pulsar electrodynamics
\cite{key-31}, in the exterior of the pulsar having an angular speed
of ${\vec{\Omega}}$, the charge density is obtained as

\begin{equation}
\rho_{e}=-\frac{\vec{\Omega}\times{\vec{B}}}{2\pi c},
\end{equation}
where $\vec{B}$ is the \emph{exterior} magnetic field. One notes
that the foregoing equation \emph{does not involve any conductivity
parameter}. Then, by drawing inspiration from this equation and by
assuming favourable boundary conditions and attendant mathematics,
one might be tempted to predict that there will be a similar charge
density in the exterior of an insulating sphere, embedded in an exterior
magnetic field, too. But, obviously, such a result would be completely
erroneous despite all nice mathematics. Why? Because, in the first
place, such a result would be based on tacit replacement of an insulator
by a conductor. How? For a conducting pulsar teeming with free electrons,
the strong induced electric field rips off electrons abundantly from
its body, and this makes the exterior too a perfect conductor ($\sigma=\infty$).
Accordingly, net Lorentz force both in the interior and exterior of
the spinning conductor could be assumed to be zero:

\begin{equation}
e\left[{\vec{E}}+\frac{\vec{v}\times{\vec{B}}}{c}\right]=0.
\end{equation}
Such a relationship becomes completely invalid for a case where $N_{e}\approx0$
and $\sigma\approx0$. Certainly, it becomes meaningless for a vacuum
BH with $\sigma\equiv0$. Thus, from the viewpoint of fundamental
physics, the assumption that an uncharged mathematical vacuum BH can
act as a NS amounts to transforming $0$ into $\infty$ (with respect
to conductivity) by means of mathematics based backed by favourable
assumptions and boundary conditions. Also, motion of rotation can
be physically defined for matter and not for pure vacuum. The mathematical
rotating BH solution may be meaningful for rotating of the central
ring singularity, but, for the rest of the vacuum Kerr BH, rotation
may be defined the way one can define rotation of a NS, a white dwarf
or any matter. However, one may imagine the motion of rotation of
a region of vacuum confined within a material boundary, say an insulating
glass shell. But BHs are not confined by any material boundary, and
even for an imagined spinning vacuum confined within a glass shell,
no induced electric current can be set up in the absence of free electrons.
Further, the rotational kinetic energy of a spinning BH must reside
with the spinning ring singularity. In order to extract its kinetic
energy, there must be an electrical circuit connecting the ring singularity
with a \textquotedbl{}load\textquotedbl{} outside the BH. But, even
assuming that induced currents are somehow set up within the vacuum
BH, the kinetic energy of the spinning ring singularity cannot be
transported away sinc \emph{nothing, not even light, escapes the event
horizon let alone the central singularity} \cite{key-34}. However,
even if we would ignore such basic physics, and, in turn, accept the
usual assumption that, in the presence of the ambient magnetic field
of the surrounding accretion disk, a vacuum BH behaves like a NS,
we must note that the ambient magnetic field of arising due to the
disk is negligible compared to what is required for being dynamically
dominant, because even at the inner edge of the disk, the magnetic
field is expected to be insignificant compared to corresponding dynamic
values. This is also so because, within the ISCO, accreting matter
is expected to experience near free fall, and, further, within the
photon sphere, the same must be under perfect radial free fall without
any turbulence and without any conversion of kinetic energy into turbulent
magnetic energy. 

Therefore, the issue of launching of ultra-relativistic jets from
supposed non-magnetic BHs suffers from twin unsolved problems: 

(i) The source of energy of the jets, as in many cases, assumed reversal
of accretion kinetic energy is insufficient to foot the power budget.

(ii) The requirement for organized large scale twisted magnetic field
lines that can confine and accelerate the jets.

Clearly, a strongly magnetized spinning BHM whose large scale magnetic
field gets twisted is much better suited to explain the origin of
relativistic jets from astrophysical BHs. Relativistic jets have been
observed from several NS binaries too; for instance Swift J0243.6+6124,
4U 0614+091, Aql X-1, Sco X-1, Her X-1, Cir X-1. 

\section{Ideas About Magnetized Black Hole Mimickers}

In 1963, Hoyle and Fowler proposed that the centre of the quasars
contain Radiation Pressure Supported Supermassive Stars and that the
luminosity of the quasars may be ascribed to the huge luminosity of
such supermassive stars radiating at their Eddington limit \cite{key-35,key-36}.
They however ignored the likely magnetized nature of such compact
supermassive stars. Also, the supermassive stars conceived by Hoyle
and Fowler were quasi-Newtonian with modest value of gravitational
compactness ($z$).

But in 1964, Ginzburg came close to conceiving of general relativistic
compact objects supported by radiation and magnetic pressures ($z\gg1$),
as he pointed out that the collapsing massive star having frozen in
magnetic field should develop strong dipole magnetic field immediately
before becoming a BH, and in fact it may end up an ultra-magnetized
``superstar'' \cite{key-3}.

In the following year 1965, Thorne showed that pure magnetic energy
would not collapse into a BH state \cite{key-37}. Essentially, he
highlighted the fact that the ultra-strong magnetic field generated
preceding BH formation, as conceived by Ginzburg, may itself inhibit
the formation of exact BHs. 

Nonetheless, the supermassive star hypothesis soon gave way to the
idea of accretion onto exact supermassive BHs as the powerhouse of
quasars. One of the reason behind this transition was that the supermassive
stars of Hoyle and Fowler were supposed to be powered by nuclear fusion
at their cores just like ordinary stars. In fact, Hoyle and Fowler
ignored the fact that even extremely show gravitational contraction
of supermassive stars might generate their luminosity despite the
absence of any central nuclear burning.

In 1969, Goldreich and Julian developed the theory of pulsars as spinning
strongly magnetized NSs where the source of energy is the rotational
kinetic energy of the NS, and neither any accretion process nor surface
luminosity of any star \cite{key-31}. Following this, in the same
year, Morrison published a paper \cite{key-38} assuming that quasars
are powered by giant pulsars, which are spinning magnetized supermassive
stars conceived by Hoyle and Fowler. Since then, despite the dominance
of the BH hypothesis, many authors extended the idea that quasars
may contain magnetized supermassive stars instead of BHs having no
magnetic field at all {[}39\textendash 54{]}. In particular, in 1977,
Ginzburg and Ozernoi \cite{key-43} coined a term ``Magnetoids''
to describe such spinning magnetized supermassive stars, \emph{supported
by radiation pressure, magnetic field and centrifugal repulsion}.
Some authors instead chose a term \textquotedbl{}spinar\textquotedbl{}
to describe such non-singular BH candidates. In 2008, Lipunov and
Gorbovskoy defined \cite{key-51}:

``A spinar is a collapsing object with quasi-equilibrium. Its equilibrium
is maintained by the balance of centrifugal and gravitational forces
and its evolution is determined by its magnetic field.''

However, all such studies are, at the best, sketches and no comprehensive
general relativistic study was ever made. In particular, such studies
did not address the crucial questions such as:

(i) How massive stars or supermassive gas clouds, during their continued
gravitational collapse, end up as non-singular compact objects when
it is commonly believed that continued gravitational collapse must
lead to formation of exact BHs?

(ii) What is the source of central energy generation of such massive
compact objects? Central nuclear burning or something else? It seems
that the Referees or Editors behind such papers published in prestigious
journals did not raise such issues either.

This theoretical vacuum was largely filled in 2000-10, and a much
more solid framework was realized for existence of quasi-static ultra-magnetic
compact objects, the so-called Magnetospheric Eternally Collapsing
Objects (MECOs), see \cite{key-6} and {[}55\textendash 61{]}. It
turned out that MECOs are essentially extremely general relativistic
versions of Radiation Pressure Supported Stars whose concept was originally
given by Hoyle and Fowler \cite{key-35,key-36}. Radiation Pressure
Supported stars are so hot that they are radiating at their Eddington
Limit where, by definition, the outward radiation pressure balances
the inward gravitational pull. 

The concept of a MECO relies on the fact that, during continued collapse,
once the massive star contracts below its \emph{photon sphere which
is a quasi-event horizon}, the heat and radiation generated by the
collapse process get trapped by self-gravity. One notes that, while
the photon sphere has $z=0.732$, the EH should have $z=\infty$.
Thus, the journey from the photon sphere upto the supposed EH should
be an infinite trek in terms of traversing through strength of gravity.

It can be seen that the outward force due to trapped radiation increases
much faster, that is $\sim(1+z_{s})^{2}$, than the relevant Eddington
luminosity $\sim(1+z_{s})$. Consequently, at some appropriately high
$z_{s}\gg1$, there should a quasi-equilibrium upon attainment of
Eddington luminosity by the collapsing object, see \cite{key-6} and
{[}57\textendash 61{]}.

While the idea of MECO eliminates the formation of exact BHs on the
strength of inevitable generation of Eddington limited radiation pressure
supported stars, it may be relevant to note that nonlinear electrodynamics
\cite{key-62} too might prevent formation of exact BHs.

\section{Astrophysical Evidences For MECOs}

For NSXBs, as mentioned earlier, the location of the inner accretion
disk depends on the NS magnetic field $B_{NS}$ and mass accretion
rate ${\dot{M}}$. Since $B_{NS}$ is fixed, $r_{i}$ varies only
with mass accretion rate $r_{i}\propto{\dot{M}}^{-2/7}$, while $r_{i}$
can keep varying for NSXBs \cite{key-23,key-24}. In contrast, for
magnetic fieldless BHCXBs, one has a \emph{fixed} value of $r_{i}=3R_{s}=6M$.
In both the cases, the radiation pressure of the X-rays emanating
from the central compact object may exert some pressure on the accretion
disk when $L_{x}\sim L_{edd}$, which is the Eddington luminosity.
While for a NSXB, one may tentatively account for modification of
$r_{i}$ due to inclusion of $L_{x}$ \cite{key-63}:

\begin{equation}
r_{i}\propto{\dot{M}}^{-2/7}~\alpha^{-1/7},
\end{equation}
where $\alpha=L_{x}/L_{edd}$. It was found that $r_{i}$ starts expanding
only if $L_{x}>0.66{\dot{M}}_{edd}$, where $M_{edd}$ is the Eddington
mass, and the effect is negligible for $L_{x}$ well below $L_{edd}$.
Thus, we may neglect the radiation pressure effect on $r_{i}$ when
$L_{x}<0.66L_{edd}$.

It is known that, for some NSXBs, $L_{x}$ may suddenly increase by
a factor of thousands for a few days. It is understood that such sudden
\emph{transient} behavior, or state, changes due to variations of
the value of $r_{i}$ because of changing ${\dot{M}}$. On the other
hand, since $r_{i}$ remains fixed for BHCXBs (except when $L_{x}\approx L_{edd}$),
one does not expect them to undergo state changes though $L_{x}$
can vary with ${\dot{M}}$. But, surprisingly, most of the BHCXBs
too are transient and that is difficult to understand. Thus, Robertson
and Leiter attempted to explain the changes of spectral states of
both NSXBs and BHCXBs occurring over periods of several years on a
common platform by assuming the BHCXBs to contain MECOs instead of
BHs \cite{key-64,key-65}.

Also, many NSXRBs and BHCXBs emit radio waves apart from X-rays. In
addition, there is a certain relationship between the radio and X-ray
luminosities of the two apparently distinct classes of binaries. Yet,
surprisingly, this relationship is practically the same for both the
cases. This suggests the existence of some intrinsic similarity between
the physical nature of variable NSXBs and BHCXBs. Robertson and Leiter
also explained this almost common relationship between the radio and
X-ray luminosities of BHCXBs and NSXBs by proposing that the former
contains magnetized compact objects (MECOs) like the latter (NS) \cite{key-66}.
They also pointed out that the quiescent weak X-ray flux from several
LMXBs, containing weakly magnetized spinning neutron stars, can be
explained ascribing the quiescent emission to the spin down luminosities
\cite{key-64}. 

We have already emphasized that the NSXBs always suffering from tug
of war between outward magnetic pressure and inward ram pressure of
the accreting plasma are expected to be BHCXBs containing magnetic
fieldless dead BHs from which nothing can escape. However, it transpires
that the BHCXBs are much more transient, much more violent than the
NSXBs. Such an apparent contradiction may be resolved by considering
that the so-called astrophysical BHs are MECOs, ultra-compact balls
of magnetized plasma, which are vulnerable to unpredictable ``flares''
and ``Coronal Mass Ejections''. It is likely that the outbursts
from the accretion disks too are triggered as they are hit by MECO
flares. There is, of course, no direct evidence for such proposals,
but, nonetheless, such a scenario looks much more reasonable than
one in which magnetic fieldless vacuum BHs can trigger ``burps''
and ``flares''. We have already mentioned that one major X-ray flare
detected from the AGN Markarian-33 might be better understood in terms
of a burst of corona from close to the BH candidate weighing $\sim2\times10^{6}M_{\odot}$.
It is tempting to assume that this was a case of CME from the MECO
of this AGN \cite{key-21}.

In 2006, astrophysicists discovered ``Fermi bubbles'', that are
two colossal elliptical gamma ray emitting blobs extending around
10 kpc above and below the GC \cite{key-67}. The blobs are filled
with very hot magnetized plasma. Even 14 years post their discovery,
``the formation mechanism of the bubbles is still elusivis'' \cite{key-68},
though there are several conjectures \cite{key-68}. Yet, the mirror-like
symmetry of the bubbles around the GC suggests that they resulted
from some super gigantic explosion at the GC which injected an energy
may be as large as $10^{55-56}\,ergs$ some $5$ to $6$ $Myr$ ago.
It is tempting to assume that it was an explosion of the MECO Sgr
A$^{*}$. Such an historic explosion may not be any Coronal Mass Ejection,
but may have been triggered by additional instability generated by
the accretion of a massive star or a small star cluster onto the MECO. 

\subsection{More Direct Evidences For MECOs}

In 2006, Schild, Leiter and Robertson presented evidence that the
inner edge of the accretion disk of the quasar QSO 0957+561, appearing
as a bright luminous ring, is located at $r_{i}\sim35R_{s}=70M$,
when, for an unmagnetized BH, one expects $r=3R_{s}=6M$. Accordingly,
they concluded that QSO 0957+561 contains a MECO in lieu of a BH.
They also inferred the existence of ``hourglass shaped'' outflow
of plasma from around the central compact object, and interpreted
this outflow to be guided by large scale organized dipole magnetic
fields of the MECO \cite{key-69}.

Later, a similar inner structure was found in another quasar Q2237
\cite{key-70}. Further, studies of 55 more quasars too indicated
the presence of inner magnetic field controlled structures, similar
to the one found in QSO 0957 \cite{key-71}. Such studies suggest
that all quasars might be containing MECOs in lieu of BHs. The detailed
physics of emission of X-ray and radio emission from Sag A$^{*}$,
the BH candidate at the centre of our galaxy, the Milky Way, may too
be well understood in the MECO paradigm \cite{key-72}. In Section
3, we have discussed that there is direct evidence for presence of
unexpectedly high large scale partially organized magnetic fields
around the compact objects of innumerable AGNs as well as for Sgr
A$^{*}$, the supposed supermassive BH in our GC {[}27\textendash 30{]}.
Further, there is evidence for a superstrong magnetic field at the
inner accretion disk of the compact object in Cygnus X-1 \cite{key-25}.
In fact, the so- compact object in Cygnus-1 may have a dipole magnetic
moment of $\sim10^{30}\,Gcm\ensuremath{^{3}}$, which led to the idea
that it is a MECO rather than a true BH \cite{key-26}. Unfortunately,
the authors of the last paper invented a new term, using the terms
\emph{Magnetic Extremely Compact Object} instead of using the term
``Magnetospheric Eternally Collapsing Object'', that has been used
in literature since 2003.

\section{Nature of the Magnetic Field Structure of MECO }

All astrophysical bodies, except Kerr and Schwarzschild BHs, possess
some intrinsic magnetic field. This magnetic field has predominantly
dipole nature. However, deviation from perfect spherical symmetry
may give rise to some quadrupole, or even higher order moments of
mass distribution. Similarly, all astrophysical bodies do possess
some quadrupole and even higher order magnetic moments too. Yet, for
analytical studies, such higher order magnetic moments are ignored
because of two reasons:

(i) Overall magnetic field is predominantly due to the dipole component.

(ii) Difficulty in handling weak higher order moments in analytical
studies.

For instance, even a rapidly spinning pulsar is treated as a spinning
magnetic dipole. This issue, in general, leads to a reasonably accurate
physical picture \cite{key-31}. Similarly, almost all analytical
studies of gravitational collapse ignore the presence of higher order
moments with respect to mass distribution or magnetic field structure
{[}74\textendash 76{]}. Two of the early models of continued gravitational
collapse had concluded that that all magnetic moments should die down
for the eventual BH \cite{key-76,key-77}. 

Finally, we know that, by the \textquotedbl{}No Hair Theorem\textquotedbl{},
at the final stages of (uncharged) BH formation all higher order magnetic
moments must vanish, and therefore, at the penultimate stage, only
dipole moment should survive. Accordingly to Ginzburg \cite{key-3}
and following him, various other authors have considered \emph{only
magnetic dipole moment for the ``magnetoids'' or ``spinars''}
{[}40\textendash 55{]}. Therefore, the dominant magnetic field of
MECO too should be given by the dipole component, that we shall consider
below. But, if in the future some other author would consider higher
order magnetic moments, however weaker, that will be welcome.

For a non-relativistic NS, the polar and radial components of the
dipole field are

\begin{equation}
B_{\theta}=\frac{\mu\sin\theta}{r^{3}}\label{eq: 5}
\end{equation}
and

\begin{equation}
B_{r}=\frac{2\mu\cos\theta}{r^{3}},\label{eq: 6}
\end{equation}
where $\mu$ is the magnetic dipole moment (as seen by a distant observer)
and $4\pi r^{2}$ denotes invariant area of symmetric 2-spheres. Here,
$B_{\theta}$ and $B_{r}$ are the components of the magnetic field
in local tetrad and $B=\sqrt{{B_{\theta}}^{2}+{B_{r}}^{2}}$. Thus,
the magnetic field strength at the pole

\begin{equation}
B_{p}=B(\theta=0)=\frac{2\mu}{R^{3}}\label{eq: 7}
\end{equation}
is twice the strength at the equatorial field
\begin{equation}
B_{e}=B(\theta=90^{o})=\frac{\mu}{R^{3}}.\label{eq: 8}
\end{equation}
Given such an asymmetry at the non-relativistic level, one might expect
that, for a general relativistic compact object, $B_{p}$ should be
larger than $B_{e}$ by a factor larger than $2.0$. However, here
we point to a previously unnoticed general relativistic effect: \emph{for
a sufficiently relativistic compact object, the polar magnetic field
becomes weaker than the equatorial field in direct contrast to their
corresponding Newtonian behaviour}. 

Yet eventually BHMs may even behave like low magnetic field millisecond
pulsars or atoll type neutron stars in X-ray binaries.

\section{General Relativistic Dipole Magnetic Field}

The general relativistic modification of assumed dipole magnetic fields
are known for many decades \cite{key-3,key-72}. The components of
field in local tetrads are:

\begin{equation}
B_{\theta}=\frac{\mu\sin\theta}{r^{3}}~F_{1}(x),\label{eq: 9}
\end{equation}

\begin{equation}
B_{r}=\frac{2\mu\cos\theta}{r^{3}}~F_{2}(x),\label{eq: 10}
\end{equation}
where $x=r/2M$, and

\begin{equation}
F_{1}(x)=6x^{3}(1-x^{-1})^{1/2}\ln{(1-x^{-1})}+6x^{2}\frac{\left(1-\frac{x^{-1}}{2}\right)}{\left(1-x^{-1}\right)^{1/2}},\label{eq: 11}
\end{equation}
\begin{equation}
F_{2}(x)=-3x^{3}\ln{(1-x^{-1})}-3x^{2}\left(1+\frac{x^{-1}}{2}\right).\label{eq: 12}
\end{equation}
One easily sees that

\begin{equation}
1+z=(1-x^{-1})^{-1/2}\label{eq: 13}
\end{equation}
and, for $z\gg1$, we have 
\begin{equation}
x=1+\epsilon;\qquad\epsilon\ll1.\label{eq: 14}
\end{equation}
Then, one sees from equations ($\ref{eq: 11}$) and ($\ref{eq: 12}$)
that, in this limit, one gets 
\begin{equation}
F_{1}(x)\approx6\sqrt{\epsilon}\ln{\epsilon}+\frac{3}{\sqrt{\epsilon}},\label{eq: 15}
\end{equation}

\begin{equation}
F_{2}(x)\approx-3\ln{\epsilon}-4.5.\label{eq: 16}
\end{equation}
One also notes that since 

\begin{equation}
\ln{\epsilon}=-2\ln{(1+z)}\approx-2\ln{z},\label{eq:17}
\end{equation}
the RHS of equation ($\ref{eq: 15}$) is dominated by the $3/\sqrt{\epsilon}$
term, while the same for equation ($\ref{eq: 16}$) is dominated by
the $\ln{z}$ term if one would consider a range of $z_{s}>10^{5}$.
While such a large value of $z_{s}$ could be surprising, one recalls
that for a true BH $z_{s}=\infty$ and that \emph{any finite number
is infinitely smaller than} $\infty$! Thus, for such extremely relativistic
BHMs, in the immediate vicinity, one has 
\begin{equation}
F_{1}(x)\sim3~z,\label{eq: 18}
\end{equation}

\begin{equation}
F_{2}(x)\sim6\ln{z}.\label{eq:19}
\end{equation}
Such analytical estimates have been verified by means of simple numerical
evaluations too:

\[
\left[\begin{array}{cccc}
Surface\,redshift\,z_{s} & 10^{10} & 10^{8} & 10^{6}\\
\\
F_{1}(x)\sim3~z_{s} & 3*10^{10} & 3*10^{8} & 3*10^{6}\\
\\
F_{2}(x)\sim6\ln{z_{s}} & 133.66 & 106.02 & 78.39.
\end{array}\right]
\]
Consequently, the magnetic fields at the equator and pole of the BHM
are respectively:

\begin{equation}
B_{e}\sim3\frac{\mu}{R^{3}}z_{s},\label{eq: 20}
\end{equation}
and

\begin{equation}
B_{p}\sim12\frac{\mu}{R^{3}}\ln{z_{s}}.\label{eq: 21}
\end{equation}
Thus, 

\begin{equation}
\frac{B_{e}}{B_{p}}\sim\frac{1}{4}\frac{z_{s}}{\ln{z_{s}}}.\label{eq: 22}
\end{equation}
Accordingly, for a BHM with $z_{s}=10^{10}$, one should expect 
\begin{equation}
\frac{B_{p}}{B_{e}}\sim10^{-8}.\label{eq: 23}
\end{equation}
Hence, even if one hypothesizes that the equatorial field of a stellar
mass BHM is stronger than that of magnetars, say $B_{e}~10^{16}\,G$,
the polar field could be very weaker: $B_{p}\sim10^{8}\,G$. Here,
one may note that, for $x\gg1$, both $F_{1}\:and\:F_{2}\sim1,$ and
the dipole field approaches the non-relativistic form

\begin{equation}
B_{r}\approx\frac{2\mu}{r^{3}}\left[1+\frac{3M}{2r}\right]\cos\theta,\label{eq: 24}
\end{equation}

\begin{equation}
B_{\theta}\approx\frac{\mu}{r^{3}}\left[1+\frac{2M}{r}\right]\sin\theta.\label{eq: 25}
\end{equation}
In any case, there will be an ambiguity about the evaluation of the
local magnetic moment at the surface of the BHM. To some extent, such
an ambiguity exists even for the non-relativistic case too.

Now, let us study the behaviour of the expected magnetic moments by
neglecting the trigonometric factors:

\begin{equation}
B_{\theta}=r^{-3}\mu F_{1}(x),\label{eq: 26}
\end{equation}

\begin{equation}
B_{r}=2r^{-3}\mu F_{2}(x).\label{eq: 27}
\end{equation}
For a moment, let us consider a BHC having $M=10M_{\odot}$ and $R=3*10^{6}\,cm$
($z_{s}\gg1$). For a perfect axisymmetric case, the observed magnetic
moment may be determined by $B_{r}.$ Thus, by using Eq. (\ref{eq: 27}),
for such a case the local value will be $\mu\sim1.3*10^{19}B_{r}\,cm^{3}.$
Therefore, by considering $B_{r}\leq10^{8}\,G,$ the magnetic moment
becomes $\sim10^{27}\,G\,cm^{3}$ or even much lower. However, for
a non-aligned rotator $B_{\theta}$ it will play its role and one
could also have $\mu>10^{27}\,G\,cm^{3}$. But it is important here
to note that magnetospheric emission takes place from the boundary
of the extended magnetosphere and not from the surface of the pulsar.
Tentatively, such a boundary may lie at the corresponding light cylinders,
that are expected to be beyond $r>6M$ or $x>3$. Since at $x=3$,
or $r=6M$, it is $F_{1}=2.74$ and $2F_{2}=2.68$, then one finds
$\mu(emission)\sim\mu(distant)$. Therefore, the non-relativistic
treatments made by Robertson and Leiter \cite{key-65,key-66} remain
approximately valid even though the basic situation is an example
of extreme general relativity. In other words, the magnetic fields
near the spinning down MECOs may very well behave like weakly magnetized
NSs away from their surfaces.

One may wonder why the ``local magnetic fields'' present in cosmic
locations matter for astronomers carrying out observations almost
infinitely far away. This is so because astronomers do not measure
the distant magnetic field by doing any local laboratory experiment.
On the other hand, they infer or measure the distant magnetic fields
as manifest in cosmic laboratories. For instance, let astronomers
detect a certain cyclotron line is a local lab. They correct the frequency
of the line by accounting for the cosmological redshifts (when due)
of the zone of generation of the line. Having done this correction,
they use the formula connecting cyclotron line frequency with the
ambient magnetic field. By this way, they measure the magnetic field
in the distant cosmic location by sitting in their own lab. In short,
astronomers do measure the local cosmic magnetic fields and not their
redshifted lab values \cite{key-30}.

\section{Discussions}

The massive compact objects found at the centre of most of the galaxies
and in many X-ray binaries are certainly not NSs. In fact, NSs cannot
be more massive than three solar masses. Such compact objects are
believed to be BHCs. Despite this, in the past two decades many authors
have suggested various alternatives to true BHs and which may generically
be termed as BHMs having radii $R\approx2M$. Also, despite the widespread
popular belief, astronomers have never detected, either indirectly
or directly any EH or any exact BH. As to the detection of the shadow
of the supposed BH in M87, it must be borne in the mind that even
for an isolated BH or BHM, what the EHT or any other telescope can
see is, at the most, the gravitationally lensed inflated shadow of
the photon sphere ($R=3M$): $R_{shadow}=\sqrt{3}~3M$. It is interesting
to note that, not only a true BH ($R=2M$) or an ideal BHM ($R\approx2M$),
but any compact object whose radius is $R\le3M$ would generate a
shadow of the same size. Thus, the so-called image of the supposed
BHC must be due to emission from the surrounding plasma beyond the
photon sphere/photon ring. 

One stresses that many astrophysical phenomena are difficult to understand
in terms of vacuum BHs that have no magnetic field and from which
nothing can escape. For instance emergence of relativistic jets from
astrophysical jets often require that BHs somehow act like magnetized
neutrons stars which can, not only launch jets from close to their
magnetized surfaces, but also keep on guiding and accelerate the jet
by its large scale magnetic fields twisted by their anchored magnetic
fields. Indeed, it is more easier to understand the emergence of relativistic
jets from magnetized NSs than from unmagnetized vacuum BHs. Accordingly,
astrophysicists are compelled to picture BHs as some sort of magnetized
NS on the plea that the exterior magnetic field lines can thread a
BH. But exterior magnetic field lines can thread an insulator too,
that does not metamorphose a moving or spinning insulator sphere into
a conducting sphere.

In fact, seen from the purely \emph{mathematical angle} of the STR,
there should be an induced electric field within any moving material
including an insolaror ${\vec{E}}_{magnetic}=\frac{1}{c}{\vec{v}}\times{\vec{B}}$
if $v\ll c$. This induced electric field becomes fructuous only in
the presence of dense free electrons when a corresponding magnetic
part of Lorentz Force is set up: ${\vec{F}}_{magnetic}=\frac{e}{c}{\vec{v}}\times{\vec{B}}$
if $v\ll c$. Since for insulators the free electron density is about
$n\approx0$, for the issue of electromagnetic induction, moving insulators
cannot behave like moving conductors even though exterior magnetic
field lines can thread both. In the same vein, a region of moving
vacuum, whatever it may mean physically, or a moving BH having $n=0,$
cannot behave as moving conductors or neutron stars.

For BH accretion, the disk magnetic field at $r_{i}$ seems too expected
to be insignificant compared to the dynamical value in order that
the accretion flow in the disk is reasonably smooth (near Keplerian)
and not disrupted. In particular, for Cygnus X-1, one would expect
a value of $B_{i}\ll10^{8}\,G$ \cite{key-22}.

In fact in 2013, Broderick and Loeb admitted that, verbatim \cite{key-78}

``More embarrassing to astrophysicists is our lack of understanding
of black hole jets - phenomena in which the forces near a supermassive
black hole somehow conspire to spew out material at ultra-relativistic
speeds (up to 99.98 percent of light speed)''.

Such a fundamental difficulty can be eliminated by realising that
the so-called astrophysical BHs could be MECO that can launch and
accelerate relativistic jets the way many NSXBs such as Cir X-1, Aql
X-1 do. There are many indirect as well as direct evidences for existence
of strong magnetic fields around so-called astrophysical BHs and that
are most naturally explained by realising that astrophysical BHs could
be MECOs.

For a NS or a MECO, the region interior to the accretion disk $r<r_{i}$,
by definition is \emph{dynamically dominated} by the magnetic field
of the central compact object. In contrast, for accretion around unmagnetized
BHs, the ambient magnetic field around the BH must be negligible because
$B_{i}$ already is dynamically negligible and near spherical free
fall of plasma cannot generate additional magnetic field. 

In NSXBs, the location of the inner accretion disk around a magnetized
NS is determined by its surface magnetic field $B_{s}$ and the mass
accretion rate ${\dot{M}}$is: 
\begin{equation}
r_{i}\propto B_{s}^{4/7}{\dot{M}}^{-2/7}
\end{equation}
For a NS spinning with an angular speed of $\omega_{s}$, one can
also define a ``corotation radius'' $r_{co}$ where the Keplerian
angular speed of the accreting plasma equals to the spin angular momentum
of the NS \cite{key-23,key-24}:

\begin{equation}
\omega_{s}=\sqrt{GM/r_{co}^{3}}.
\end{equation}
In the NS magnetosphere, plasma can corotate with the NS only up to
$r\le r_{co}$, and it transpires that steady accretion is possible
only for

\begin{equation}
r_{i}<r_{co};~~steady~accretion,
\end{equation}
a state which happens for high mass accretion rates (${\dot{M}}$).
But the plasma at $r>r_{co}$ cannot corotate with the NS and the
accreting material get ejected or ``propelled'' out for 
\begin{equation}
r_{i}>r_{co}~ejection~or~propeller~regime,
\end{equation}
which is a state which happens for very low mass accretion rates (${\dot{M}}$).

State transitions in NSXBs are usually explained in terms of the above
scheme, with significant variation of the mass accretion rate. But,
in contrast, for BH accretion there is \emph{no magnetosphere,} no
concept of any ``corotating radius'', and a fixed value of $r_{=}3M$.
Thus, it is difficult to explain the origin of state transitions in
BHCXBs. It is here that Robertson and Leiter could explain the observed
state transitions in BHCXBs by invoking magnetized MECOs in lieu of
unmagnetized BHs \cite{key-65,key-66}. Similarly, they could explain
the origin of almost similar Radio - X-ray correlations in both NSXBs
and BHCXBs by invoking magnetized compact objects in both the cases.
Such exercises suggest that the so-called astrophysical BHs may be
MECOs \cite{key-66}.

Schild, Leiter and Robertson found that the inner edge of the accretion
disk of the quasar QSO 0957+561, appearing as a bright luminous ring,
is located at $r_{i}\sim35R_{s}=70M$, when for an unmagnetized BH,
one expects $r_{=}3R_{s}=3M$ \cite{key-69}. This observation provided
more direct evidence that the so-called BHs may be MECOs. The hourglass
shaped outflow from the central region of this quasar gets naturally
explained by the large scale organized dipole magnetic field of the
MECO \cite{key-69}. Similar hourglass shaped outflow has been detected
from many more quasars suggesting that, in general, quasars contain
MECO \cite{key-70,key-71}.

As to direct evidence, it was found that, for this source, there is
a strong magnetic field $B\approx10^{8}\,G$ at the inner accretion
disk of Cygnus X-1 \cite{key-25}, when the same is expected to be
$B\ll10^{8}\,G$ \cite{key-22}. Thus, Cygnus-1, which apparently
confirmed the existence of astrophysical BHs in the 1970s, may be
a MECO in lieu of a BH \cite{key-26}.

In recent years, it has also been found evidence for the existence
of unusually strong large scapartially organized magnetic fields around
the supposed supermassive BH in our galaxy as well as in many AGNs
{[}27\textendash 30{]}.

In addition, it has been recently suggested that one should use the
magnetic field pattern around accreting compact objects to test whether
they are true BHs or not. While uncharged BHs possessing EHs do not
possess any magnetic field, the plasma accreting onto them can possess
magnetic fields and which may vary as $\sim r^{-1}$ \cite{key-8}.
In contrast, even uncharged BHMs may possess their own magnetic fields,
and in particular, MECOs should possess strong intrinsic magnetic
fields. At first sight, the magnetic field around a magnetized BHM
should be dominated by the intrinsic dipole field falling off as $r^{-3},$
because the magnetic field in the plasma ($\sim r^{-1}$) is expected
to be much weaker to the intrinsic field of the BHM \cite{key-8}.

Here we highlight the fact that, in the immediate vicinity of the
BHM, the field pattern should be much more complex than the one of
a NS: $B\sim r^{-3}$. This is because while the radial field is $B_{r}\sim3\frac{\mu}{R^{3}}z_{s}$,
the polar field is $B_{p}\sim12\frac{\mu}{R^{3}}\ln{z_{s}}$. Thus,
on the surface of the BHM, the magnetic field at the pole ($B_{p}$)
is weaker than the same at equator ($B_{e}$) by an extremely large
factor of $\sim z_{s}/lnz_{s}$. Therefore, even an ultra magnetized
BHM may behave as a NS whose magnetic field could be weaker than the
same of young NSs by a factor of order $10^{4}$. 

\section{Concluding Remarks}

Astronomers have not detected any EH, and, by definition, it is fundamentally
impossible to directly detect the same \cite{key-2}. There is no
proper explanation for innumerable indirect or direct evidence for
unexpected strong magnetic fields around astrophysical BHs. In particular,
the issue of detection of an unusually strong magnetic field in the
vicinity of the Sgr A$^{*}$, the supposed supermassive BH at the
galactic centre, has recently been scrutinized by Peng, Liu and Chou
\cite{key-79}. They concluded that:

``the observed ultra-strong radial magnetic field near the GC can
not be generated by the generalized $\alpha$-turbulence type dynamo
mechanism since preliminary qualitative estimate in terms of this
mechanism gives a magnetic field strength \emph{six orders of magnitude
smaller} than the observed field strength at $r=0.12$ pc''. Such
a conclusion is in perfect agreement with the conclusion of Shakura
and Sunayev almost half a century ago \cite{key-22}. In order to
resolve this problem, these authors proposed that Sgr A$^{*}$ is
a super-massive star with magnetic monopoles. However, such a suggestion
has hardly any physical basis as there are no magnetic monopoles.
On the other hand, recent theoretical research on general relativistic
astrophysics has shown that once a massive star or a supermassive
gas cloud will dip below its ``photon sphere'' ($z=0.723$) during
their continued collapse, sooner or later ($z\gg1$), the luminosity
of the radiation trapped by their own gravity must attain the Eddington
Limit when the collapsing star or the gas cloud would turn into a
quasi-static ultra-hot ball of plasma whose energy density is dominated
by radiation. In the absence of any EH ($z=\infty$), the quasi-static
radiation ball must keep on ECO towards the $z=\infty$ exact BH state,
see \cite{key-6} and {[}56\textendash 62{]}. The magnetic field frozen
in astrophysical plasma increases as $B\sim R^{-2}$ and further there
will be general relativistic enhancement at high $z\gg1$, a phenomenon
which was stressed by Ginzburg way back in 1964 \cite{key-3}. Thus,
ECOs must have a strong local magnetic field and a surrounding magnetosphere.
Hence, ECOs must be Magnetospheric ECOs (MECOs). To a good approximation,
the magnetic field of all astrophysical objects is dominated by the
dipole component. Then, since all higher order magnetic fields must
be radiated away one by one before formation of exact (uncharged)
BHs, MECOs could be almost perfect magnetic dipoles as first envisaged
by Ginzburg in 1964 \cite{key-3}. Therefore, evidence for the existence
of unexplained strong partially organized magnetic fields can be best
explained by the realisation that the so-called astrophysical BHs
may be MECOs. Furthermore, in view of the claim of the presence of
GW echoes in the LIGO signals, the question of differentiating a true
BH with a BHM becomes very pertinent. One tool to distinguish a true
accreting astrophysical BH from an ultra-magnetized BHM could be through
study of the magnetic field structure around the relevant compact
object. For a true BH, one may expect the frozen in plasma magnetic
field to fall off as $B\sim r^{-1}$. On the other hand, for a magnetized
BHM, while far away from the surface $B\sim r^{-3}$, in the immediate
vicinity, the field pattern is dramatically asymmetric and complex:
the radial field $B_{r}\sim3\frac{\mu}{r^{3}}z_{s}$ and the polar
field $B_{\theta}\sim12\frac{\mu}{r^{3}}\ln{z_{s}}$. In fact, radial
field pattern apart, any inference about the dramatic asymmetry of
the equator field $B_{e}$ and polar field $B_{p}$ components may
suggest existence of magnetized BHMs or \emph{magnetoids} conceived
by Ginzburg \cite{key-3}, or giant pulsars conceived by Morrison
\cite{key-38}, or MECOs, or any other magnetized BHMs.

Higher magnetic fields, if present, must be extremely weaker in comparison
to the dominant dipole component. All studies of general relativistic
gravitational collapse of a magnetized star consider only the dipole
magnetic moment. But it would be welcome if in future, some studies
would consider higher order magnetic moments in this context. Irrespective of such details, in the ultimate analysis, the so-
called astrophysical black holes cannot be true black holes for the following reason. The BH paradigm presumes that
mathematical BHs necessarily have $M > 0$ and the event
horizon (earlier known as Schwarzschild Singularity) must
be a sphere. But way back in 1969, Bel showed that the event
horizon actually behaves as a point singularity \cite{key-80} implying that
the latter corresponds to M = 0 and not M > 0 Bel (1969).

The same result has been independently obtained from vari-
ous other analyses too \cite{key-81, key-82}. Hence it is indeed
highly likely that the astrophysical BH candidates are only
BH mimickers and possess strong intrinsic magnetic fields
unlike mathematical BHs.

\section{Acknowledgements }

The Authors are thankful to the anonymous referee for his/her constructive
critique and seeking clarification of many issues. We accordingly
almost rewrote the manuscript and we believe this present version
has much more scientific value than the original one.

\end{document}